# Survey on the Event Orderings Semantics Used for Distributed System

Yaser Miaji  Osman Gazali and Suhaidi Hassan

ymiaji@gmail.com     osman@uum.edu.my     Suhaidi@ieee.org

UUM College of Arts and Sciences
University Utara Malaysia
06010, UUM Sintok, Malaysia
ymiaji@gmail.com

***Abstract***—Event ordering in distributed system (DS) is disputable and proactive subject in DS particularly with the emergence of multimedia synchronization. According to the literature, different type of event ordering is used for different DS mode such as asynchronous or synchronous. Recently, there are several novel implementation of these types introduced to fulfill the demand for establishing a certain order according to a specific criterion in DS with lighter complexity.

**Purpose –** This paper demonstrates most significant implementation of types of event ordering in DS.

**Designing, methodology, approach –** This paper firstly, present each type separately. Then it presents its implementation approaches. The comparison between those types is achieved later in the paper.

**Finding –** Most types used in event ordering in DS share same properties with some delicate differences. However, some types which used for asynchronous mode cannot be used in the synchronous.

**Value –** This paper is considered as a reference for scholars how desire to direct their research; develop potential investigation; or introduce new type of event orderings in DS.

*Keywords-component;*

Distributed system; event ordering;

## 1. INTRODUCTION

A distributed system refers to a system that consists of a number of computers that do not share a memory or a clock and communicate with each other by exchanging messages over a communication network. A distributed operating system operates on multiple autonomous computers but appears to its users as a single machine. Figure (1-1) illustrates the embedded and basic architecture and features of distributed system.

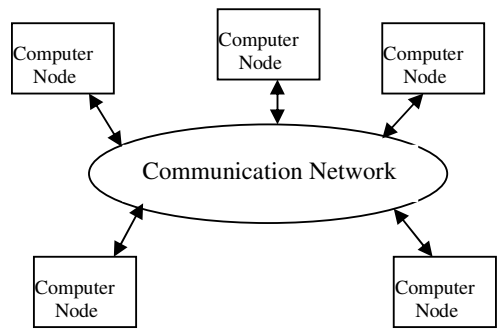

Fig (1-1): Architecture of a Distributed System

Obviously, there are several benefits from using such system. Performance is improved and the cost is reduced with the exception of certain special computation intensive applications, equivalent computing power may be obtained with a network of workstations at a much lower cost than a traditional time-sharing system. Requests of services may be satisfied using hardware/software resources on other computers on the communication network which means an enormous increase in resource sharing [4]. Furthermore, concurrent execution of tasks and load distributing can lead to improved response time. Moreover, fault tolerance can be achieved through the replication of data and services. Finally, one distinct feature is new hardware and software resources can be added without replacing the existing resources which called as modular expandability [8].

On the other hand, there are several drawbacks of using distributed system as highlighted next. The global knowledge problem which means that global state of the system is hard to acquire due to the unavailability of a global memory and a global clock and the unpredictability of message delays [10]. Naming is another problem which means the directory of all the named objects in the system (services, files, users, printers, etc.) must be maintained to allow proper access. Both schemes of replicated directories and partitioned directories have their strengths and weaknesses. Scalability which defined as any mechanisms or approaches adopted in a system must not result in badly degraded performance when the system grows. Compatibility which means that the interoperability among the resources in a system must be an integral part of the design of a distributed system is another issue. Also, process synchronization is especially difficult in distributed systems due to the lack of shared memory and a global clock [5]. Resource management refers to schemes and methods devised to make local and remote resources available to users in an effective and transparent manner. Moreover, securities which cope with two issues are relevant: authentication (verifying claims) & authorization (deciding and authorizing the proper amount of privileges). Structuring which defines how various parts of the operating system are organized [7].

On the top of all these issues appears the lack of common memory a system wide common clock is an inherent problem in distributed systems [14]. In the absence of global time, it becomes difficult to talk about temporal order of events. Without a shared memory, an up-to-date information about the state of the system is not available to every process via a simple memory lookup. The state information must therefore be collected through communication. The combination of unpredictable communication delays and the lack of global time in a distributed system make it difficult to know how up-to-date collected state information really is [1].

After the demonstration of a brief pros and cons of distributed system the reminder of the paper is organized as following; next section present the importance of event ordering in distributed system. The most popular event ordering scheme is presented in section four. The implementation of these schemes is presented in section five [3]. Finally, the comparison and conclusion are drawn.

## 2. THE IMPORTANCE OF EVENT ORDERING

Event ordering in distributed system consists in establishing a certain order among the events that occur according to some particular criteria. According to the chosen criteria, the resulting event ordering allows a greater of smaller degree of asynchronous execution. In a distributed system, there are three of events; internal, send and receive events. The internal events occur inside a process and they are never known by the rest of the participants. On the other hand, the send and receive events are those through which the participants communicate and cooperate. In this study, it is only considered the send and receive events since they modify the global state of a system [2].

There are several problem associated with event ordering in DS. For example, if there is event a and event b, how definite someone would say that a occur before b particularly in the absent of physical clock such the DS case. From this point of view the demand for a Symantec to organize and order the events is appeared [9]. There are two broad categories of events ordering; total ordering and partial ordering. Beside the ordering scheme there are two more methods used in the ordering which is no-ordering and Firs Come Firs Serve (FCFS). However, in order to define any ordering scheme there should be a definition of which event occur first [7]. In this regards, there are many approaches which is presented in next section. The implementation of each of the ordering category which varies and hence the research toward the improvement of the implementation is potential, is presented latter. Next section presents most significant event ordering scheme in more details.

## 3. SCHEME USED FOR DEFINING THE ORDER OF EVENTS

As stated formerly, there are two broad scheme; the total and partial event orderings. Before establishing the demonstration of the event ordering scheme, there will by a presentation of how the order is defined. In other words which event occurs before the other particularly in absence of physical time figure (3-1) is used to aid the explanation.

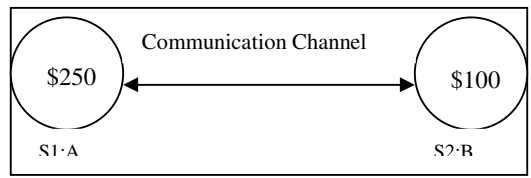

Figure (3-1): A DS with two computer nodes

### 3.1. Hppened befor relationship

In this scheme there are three rules that define that a is occurred before b:

1. If a and b in the same process and a occurred before b.
2. If a is a sending process and b is a receiving for the same process.
3. If a occurred before b and b occurred before c then a is occurred before c.

To sum up, the happened-before relation captures the causal dependencies between events, for instance whether two events are causally related or not. Event a causally affects event b if a → b. Two events a and b are said to be concurrent (denoted as allb) if not (a → b or b → a). In other words, concurrent events do not causally affect each other.

### 3.2. Logical Clock

A clock $C_i$ is associated with each process $P_i$ in the system, that can be thought of as a function for assigning a number $C_i(a)$ to any event $a$, called the *timestamp* of event $a$, at $P_i$. The happened before relation "→" can now be realized by using the logical clocks if the following conditions are met:

[**C1**] For any two events $a$ and $b$ in a process $P_i$, if $a$ occurs before $b$, then $C_i(a) < C_i(b)$.

[**C2**] If $a$ is the event of sending a message $m$ in process $P_i$ and $b$ is the event of receiving the same message $m$ at process $P_j$, then $C_i(a) < C_j(b)$.

These two conditions are guaranteed with the following implementation rules:

[**IR1**] Clock $C_i$ is incremented between any two successive events in process $P_i$ as follows:

$C_i := C_i + d$, where $d > 0$.

[**IR2**] If event $a$ is the sending of message $m$ in process $P_i$, then message $m$ is assigned a timestamp $t_m = C_i(a)$. On receiving the same message $m$ by process $P_j$, $C_j$ is first set using [IR1], then set to a value greater than or equal to the new $C_j$ and greater than $t_m$, i.e., $C_j := \max(C_j, t_m + d)$, where $d > 0$.

With these two implementation rules, two causally related events $a$ and $b$ such that $a \rightarrow b$ will have $C(a) < C(b)$, and two successive events $a$ and $b$ in process $P_i$ will yield $C_i(b) = C_i(a) + d$. (See Figure 5.3 for an example of Lamport's logical clocks) [3].

Lamport's system of logical clocks implements an approximation to global/physical time that is referred to as *virtual time*. The virtual time advances along with the progression of events and is therefore discrete. The virtual time is defined based on an irreflexive partial order "$\rightarrow$", and can be used to totally order events in a distributed system (hence produces a total order relation "$\Rightarrow$") as follows:

If $a$ is any event at process $P_i$ and $b$ is any event at process $P_j$, then $a \Rightarrow b$ if and only if either

$C_i(a) < C_j(b)$, or

$C_i(a) = C_j(b)$ and $P_i \, p \, P_j$

where p is any arbitrary relation that totally orders the processes to break ties (e.g., process id i < j implies $P_i \, p \, P_j$).

It should be noted that $a \Rightarrow b$ does not necessarily imply $a \rightarrow b$. And this is known to be a major limitation of Lamport's [15] logical clocks: If $a \rightarrow b$ then $C(a) < C(b)$, but the converse is not necessarily true. Figure 5.4 illustrates this limitation.

### 3.3. Vector Clock

Each process $P_i$ in a distributed system with $n$ processes is equipped with a vector clock $C_i$. The clock $C_i$ consists of an integer vector of length $n$, and can be viewed as a function that assigns a vector $C_i(a)$ to any event $a$ at Pi as the event's timestamp. $C_i[i]$, the $i$th entry of $C_i$, corresponds to $P_i$'s own logical time. $C_i[j]$, $j \neq i$, indicates the time of occurrence of the last event at $P_j$ that "happened before" the current point in time at $P_i$ [12]. It therefore represents $P_i$'s best guess of the logical time at $P_j$, and must satisfy the assertion of $C_i[j] \leq C_j[j]$.

The vector clocks can be implemented with the following implementation rules:

[**IR1**] Clock $C_i$ is incremented between any two successive events in process $P_i$ as follows:

$C_i[i] := C_i[i] + d$, where $d > 0$.

[**IR2**] If event $a$ is the sending of message $m$ in process $P_i$, then message $m$ is assigned a timestamp $t_m = C_i(a)$. On receiving the same message $m$ by process $P_j$, $C_j[j]$ is first incremented as in [IR1], then $C_j$ is updated as follows:

$\forall k, C_j[k] := \max(C_j[k], t_m[k])$.

Figure 5.5 shows examples of how vector clocks advance as events occur.

With vector clocks, $a \rightarrow b$ iff $t^a < t^b$, where $t^a$ and $t^b$ denote the vector timestamps of events $a$ and $b$, respectively. In other words, vector clocks allow us to order events in a distributed system and decide whether two events are causally related based simply on the timestamps of the events. The next section shows an application of vector clocks in causal ordering of messages.

## 4. EVENT ORDERING SEMANTICS

Under the two major categories of event ordering schemes, namely: total and partial event ordering in DS, there are six famous ordering schemes as following; FIFO, causal ordering, total ordering, $\Delta$-causal ordering, causal total ordering, fuzzy causal ordering and partial ordering.

### 4.1. First In First Out (FIFO)

This is the simplest ordering scheme which requires no mathematical overhead. Its rule is quite simple; if process sent event m before event n then no process send n before m. However, such scheme could cause a problem if the computer node desperately demands for a specific order for a potential reason.

### 4.2. Causal Ordering (CO)

Causal ordering of messages refers to the preservation of causal relationship that holds among "message send" events in the corresponding "message receive" events. That is, $Send(M_1) \rightarrow Send(M_2)$ implies $Receive(M_1) \rightarrow Receive(M_2)$, where $Send(M)$ and $Receive(M)$ represent the event of sending and receiving message $M$, respectively. Causal ordering of messages is important in some applications, e.g., replicated database systems, where every process in charge of updating a replica receives the updates in the same order to maintain the consistency of the database. Causal ordering of messages is not automatically guaranteed in distributed systems, hence will require implementation where necessary [5].

This scheme could be simplified as following; if sending message m causally precedes sending message m' then no process delivers m' before delivering m. However, the implementation of such scheme is not that simple. There are many proposed implementation in the area of causal ordering and research in this area is active [5].

In a causally ordered network, when a node receives a message, before the node can respond to the message it must be certain that it will not receive any other message from any other node that causally precede that message. Therefore, the node must utilize some type of ordering scheme to signify when it can respond [14].

CO is at the communication level, but consistency requirements are typically expressed in terms of the application's state. CO is not adequate in itself to ensure application-level consistency, and providing additional mechanism at the state level to remedy this deficiency eliminates the need for CO, or it is expensive. CO provides atomicity by buffering messages but fail to provide durability to message delivery [4].

Causal relationships can arise between messages at the semantic level that are not recognizable by the happens-before relationship on messages. Causal Ordering can be preserved at shared resource level but not at communication level [11]. Furthermore, CO cannot ensure serializable ordering between operations that correspond to groups of messages. Many semantic ordering constraints are not expressible in the happens-before relationship, and hence not enforceable by CO. Such ordering constraints, include causal memory, linearizability and serializability.

### 4.3. Total ordering

Total ordering semantic implies that all messages are reliably delivered in sequence to all members of a group. Also, total ordered semantic guarantees that all group members see the same order of messages. All messages arriving at al workstations are ordered [15]. Total ordering is the most stringent ordering as all message transfers between all members of the group are in order. This implies that all processes within the group perceive the same total ordering of messages. In causal ordering we are concerned with the relationship of two messages while in total ordering we are concerned with seeing the same order of messages for all group member processes [7].

Total ordering insures that each correct member delivers all messages in the same relative order. Of course, the total ordering must not violate the causal ordering, since the property of total ordering is stronger than causal ordering.

To simplify this scheme, a simple argument is provided. If correct processes p and q delivers m and m' then p delivers m before m' only if q delivers m before m'.

To place a total ordering on the set of all system events, systems of clocks could be used to satisfy the clock condition. Imply all events are ordered according to which they occur. Firstly, and y arbitrary total ordering used. And the following relationship is defined: if z is an event in process $P_i$ and b is an event in process $P_j$, then a b if and only if either (i) $C_i(a)$ or (ii) $C_i(b)$ an $P_i \rightarrow P_j$. In other words the total ordering is a way of completing the happened before [13].

The total ordering of requests leads to ineffeciency due to more data movement and synchronization requirements than what a program may really call for.

### 4.4.    Δ-Causal Ordering

The main purpose for developing Delta causality order is for some distributed applications which have to be delivered according to casual ordering and have a limited lifetime after which their data can no longer be used by the application. The first development of such scheme is by Fidge [12]. In delta scheme, the system strives to deliver as many messages as possible before their deadlines in such a way that these deliveries respect causal order [2].

The implementation of this semantic suffers from several drawbacks. It suffers from the typical pitfall of the time stamping (logical or physical) technique; to ensure causal order, in the context of broadcasting, messages have to carry a vector of integers whose dimension is given by the number of process which eventually, introduce an extra over head and complexity [4].

### 4.5.    Causal Total

This semantic is applicable if the messages satisfy the causal and the total constrain. For example, if event a occurred before event b and a is the cause for event b it partially satisfied the causal scheme. Furthermore, if both messages could be delivered to all nodes in the same order this will satisfy the total order constrains. Therefore, this message could be causal total [4].

### 4.6.    Fuzzy causal order (FCO)

The fuzzy causal relationship could be defines as following; A causally increases B" means that if A increases then B increases and if A decreases then B decreases. On the other hand if 'A causally increases B" means that if A increases then B decreases and if A decreases then B increases [1].

In the concepts that constitute causal relationship, there must exist a quantitative element that can increase or decrease.

FCM fuzzy relations mean fuzzy causality. Causality can have a negative sign. The negative fuzzy relation between two concept nodes is the degree of relation with "negation" of a concept node. For example, if the concept node $C_i$ is noted as $C_j$ the $R(C_i,C_j)=-0.6$ which means that $R(C_i,\sim C_j)=0.6$ conversely $R(C_i,C_j)=0.6$ the $R(C_i,\sim C_j)=-0.6$.

There are some principles for the FCO [1]. Firstly, If two causal relationships support the same conclusion, then the addition of those 2 causality value is > each causality value. Secondly, If a causal relationship is connected consecutively to a causal relationship, then the absolute value of its additive value of the 2 causality values is <= the least of absolute value of the 2 causality. Thirdly, The final additive value remains same irrespective of the order of addition of causality values of interest. Finally, The final causality value lies in the interval [-1,1]. However, all these principles and ties provide additional complexity and hence inefficiency to the scheme.

### 4.7.        Partially ordering

The definition of partially ordering concurred with the happened before which presented earlier. In the absence of real clock it hard to define which message is occur first. Even with the presence of real clock it is hard to accurately adjust the clock particularly with micro seconds. Therefore, the delivery of the messages in the semantic is partially ordered.

Partially ordering begins with a precise definition of the system. The assumption of a system with a collection of processes is put forward. Each process consists of sequence of events. The execution of the event should be one event [8].

From this point on, the entire system is considered as a sequence of processes [12].

For example, if a and b are events in the same process and a comes before b, then a $\rightarrow$ b. Also, if a is the sending of messages by on process and b is the receipt of the same message by another process, then a $\rightarrow$b. Finally, if a $\rightarrow$ b and b $\rightarrow$ c then a $\rightarrow$ c. Two distinct events a and b are said to concurrent then a $\rightarrow$b and b $\rightarrow$a.

### 5.  CONCLUSION

In this paper seven ordering semantic has been presented. Some of the benefit and drawbacks of using these schemes is demonstrated. This paper is a comprehensive investigation in the event ordering schemes in the distributed system. It appears that most of the semantics are concurred in several properties with a slight difference. The obvious issue is the light and simple implementation of some efficient event ordering scheme such as FCO.

### Acknowledgements

Author would like to thank his wife Ashwag for her support.

**Authors**

**Eng. Yaser Miaji**

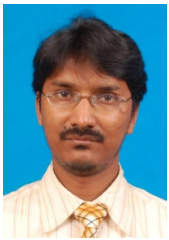

Doctoral researcher in Information Technology College of Arts and Sciences, University Utara Malaysia

Master in Telecommunication Engineering University of New South Wales, Australia, 2007

Bachelor in Electrical Engineering, Technical College in Riyadh, Saudi Arabia, 1996

Lecturer in College Of Telecommunication and Electronics in Jeddah,

Member of IEEE and ACM

**Dr. Osman Ghazali**

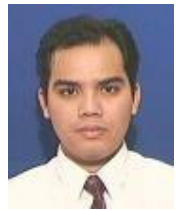

Senior lecturer graduate department of computer science, college of arts and sciences

Phd of information technology (computer network), University Utara Malaysia, 2008.

Master science of information technology, University Utara Malaysia, 1996.

Bachelor of information technology, University Utara Malaysia, 1994

**Dr. Suhaidi Hassan**

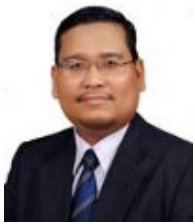

Associate Professor & Assistant Vice Chancellor at University Utara Malaysia

PhD degree in Computing (specializing in Networks Performance Engineering) from the University of Leeds in the United Kingdom.

MS degree in Information Science (with concentration in Telecommunications and Networks) from the university of Pittshugh, USA.

BSc degree in Computer Science from Binghamton University, USA.

Senior Member of IEEE